\documentclass[twocolumn]{aastex62}
\usepackage{amsmath}
\graphicspath{{./}{figures/}}

\shorttitle{Risks for Life on Proxima b}
\shortauthors{Siraj \& Loeb}

\begin{document}

\title{Risks for Life on Proxima b from Sterilizing Impacts}

\email{amir.siraj@cfa.harvard.edu, aloeb@cfa.harvard.edu}

\author{Amir Siraj}
\affil{Department of Astronomy, Harvard University, 60 Garden Street, Cambridge, MA 02138, USA}

\author{Abraham Loeb}
\affiliation{Department of Astronomy, Harvard University, 60 Garden Street, Cambridge, MA 02138, USA}



\begin{abstract}
We consider the implications that a debris belt located between Proxima b and Proxima c would pose for the rate of large asteroid impacts that could sterilize Proxima b from life. Future observations by ALMA or JWST could constrain the existence of an asteroid belt in the life-threatening regime. We generalize our rate calculation of sterilizing impacts for habitable planets in systems with an asteroid belt and an outer planet. 

\end{abstract}

\keywords{M dwarf stars -- Habitable planets -- Debris disks -- Asteroid belt -- Impact phenomena}


\section{Introduction}
Proxima b is an Earth-mass planet in the habitable zone of the nearest star, Proxima Centauri ($M_{\star} = 0.12 \mathrm{\; M_{\odot}}$), at a separation of $\sim 0.05 \mathrm{\; AU}$ \citep{2016Natur.536..437A}. Proxima b is thought to possibly hold potential for life \citep{2016A&A...596A.111R, 2016A&A...596A.112T}. Proxima c, an outer planet orbiting at a distance of $\sim 1.5 \mathrm{\; AU}$ with a mass of $\sim 10 \; M_{\oplus}$, was recently discovered \citep{2020SciA....6.7467D, 2020A&A...635L..14K}. A warm dust belt with a total mass of $\sim 10^{-3} \; M_{\oplus}$ at a distance of $\sim 0.4 \; \mathrm{AU}$ from Proxima Centauri was reported \citep{2017ApJ...850L...6A}, but later disputed as a possible stellar flare \citep{2018ApJ...855L...2M}. A possible connection between flare activity and planitesimal accretion has also been considered \citep{2011Obs...131..212B}.

In the Solar system, Saturn sets the $\nu_6$ secular resonance \citep{2006AdSpR..38..817I, 2011ApJ...732...53M}, which controls the inner edge of the asteroid belt and therefore the rate of impacts from near-Earth asteroids \citep{1994A&A...282..955M, 2000Icar..145..301B}. The relation between Saturn's location relative to the asteroid belt and mass and the impact rate have been explored through numerical simulation \citep{2018MNRAS.473..295S}. If an asteroid belt exists between Proxima b and Proxima c, Proxima c could control the rate of asteroid impacts on Proxima b. Here, we consider the risks that an asteroid belt located between Proxima b and Proxima c would pose for life on Proxima b. While asteroid impacts can help foster conditions for life (see \citealt{2018arXiv180105061L} for a review), sufficiently large impacts can boil off oceans and devastate prospects for both the development and survival of life \citep{1988Natur.331..612M, 1989Natur.342..139S, 2009Natur.459..419A, 2017NatSR...7.5419S}.




Our discussion is structured as follows. In Section \ref{sec:eir}, we consider the rate of sterilizing impacts on Earth from the asteroid belt owing to Saturn. In Section \ref{sec:pbsr}, we apply a similar calculation to Proxima b, given the existence of Proxima c. In Section \ref{sec:ire}, we investigate the generalized sterilizing impact rate for habitable worlds in multiplanetary systems. In Section \ref{sec:jwst}, we evaluate the detectability of an asteroid belt between Proxima b and c with JWST and ALMA. Finally, in Section \ref{sec:d} we explore key predictions and implications of our model.


\section{Earth impact rate}
\label{sec:eir}

For an asteroid impact on Earth, the final crater diameter $D_{\mathrm{cr}}$ is related to the impactor diameter $D_{\mathrm{imp}}$ as follows \citep{2005M&PS...40..817C},

\begin{equation}
\begin{aligned}
\label{cratersize}
    \left(\frac{D_{\mathrm{cr}}}{\mathrm{km}}\right) \sim 29 \;   & \left(\frac{D_{\mathrm{imp}}}{\mathrm{km}}\right)^{0.78} \; \left(\frac{v_{\mathrm{imp}}}{20 \mathrm{\; km \; s^{-1}}} \right)^{0.44} 
    \\ & \left(\frac{\rho_{\mathrm{imp}}}{\rho_{\oplus}} \right)^{1/3} \left(\sin{\theta} \right)^{1/3} \; \; ,
\end{aligned}
\end{equation}
where $v_{\mathrm{imp}}$ is the impact speed, $\rho_{\mathrm{imp}} \sim 2 \mathrm{\; g \; cm^{-3}}$ is the impactor density, $\rho_{\oplus} \sim 3.5 \mathrm{\; g \; cm^{-3}}$ is the density of the Earth's mantle, and $\theta$ is the angle of the impact with respect to the surface of the Earth.

The observed cratering rate on Earth is \citep{2015E&PSL.425..187H},

\begin{equation}
\label{earthrate}
    \Gamma_{\oplus} \approx 6.64 \times 10^4 \; \mathrm{Gyr^{-1}} \; \left(\frac{D_{\mathrm{cr}}}{\mathrm{km}}\right)^{-2.557} \; \; ,
\end{equation}
which is thought to be complete for the largest crater sizes considered. The time-dependence is unimportant for the purposes of this analysis as the timescales considered here are $\gtrsim 1 \; \mathrm{Gyr}$.

\begin{figure}
  \centering
  \includegraphics[width=1\linewidth]{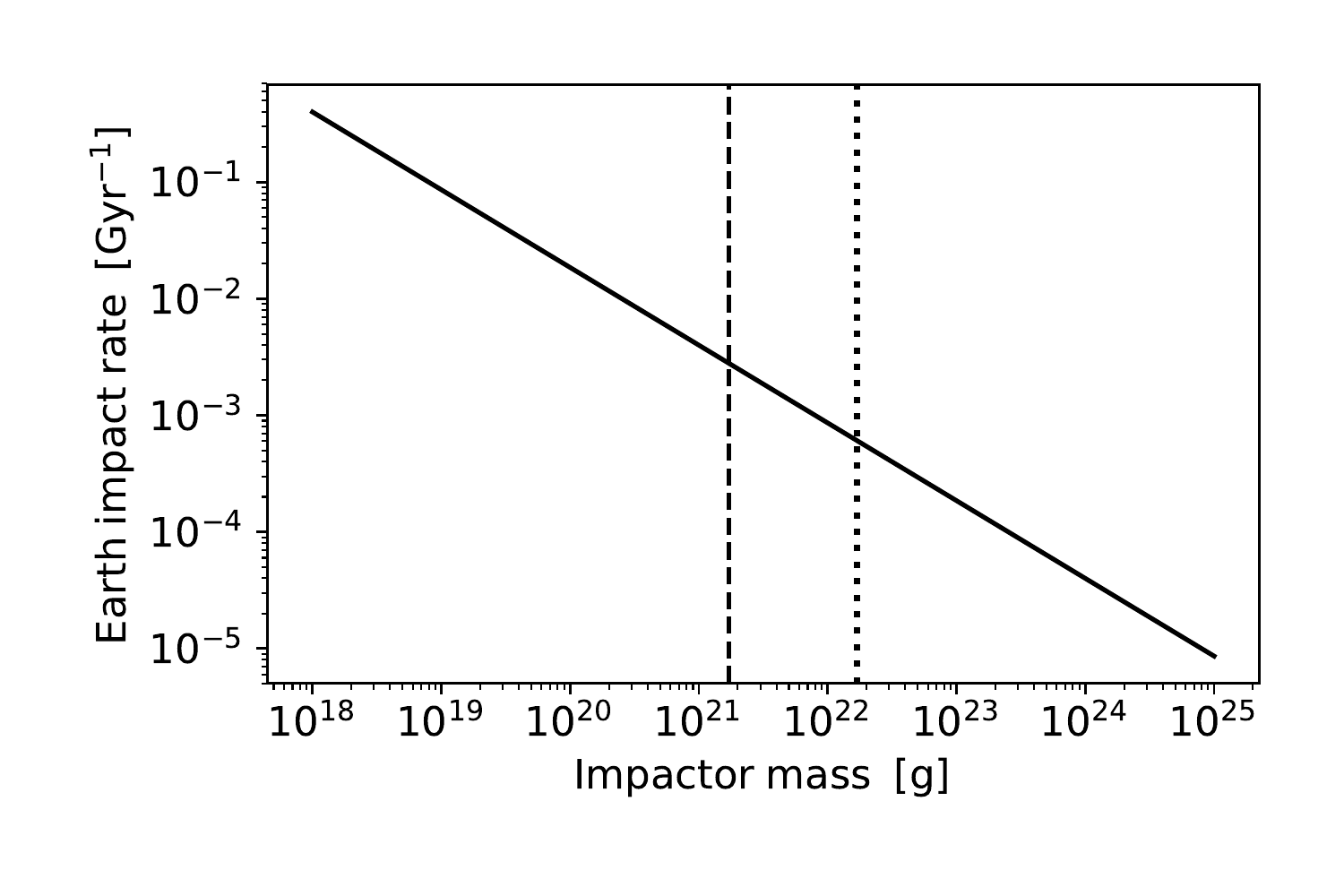}
    \caption{Asteroid impact rate on Earth (solid line) in $\mathrm{Gyr^{-1}}$ as a function of impactor mass in $\mathrm{g}$, with the dotted line indicating an impact with enough energy to boil off all of the oceans on Earth, in which 100\% of the kinetic energy is converted into thermal energy.
}
    \label{fig:earth}
\end{figure}

Substituting $D_{\mathrm{imp}} \sim (6M_{\mathrm{imp}} / \pi \rho_{\mathrm{imp}})^{1/3}$ into Eq. \eqref{cratersize} and subsequently into Eq. \eqref{earthrate} yields,

\begin{equation}
\label{finalearthrate}
\begin{aligned}
    \Gamma_{\oplus} \approx 8 \times 10^{-4} \; \mathrm{Gyr^{-1}} \; & \left(\frac{M_{\mathrm{imp}}}{\mathrm{1.7
    \times 10^{22} \; g}}\right)^{-2/3} \left(\frac{\rho_{\mathrm{imp}}}{2 \mathrm{\; g \; cm^{-3}}} \right)^{0.38}
    \\ & \left(\frac{v_{\mathrm{imp}}}{20 \mathrm{\; km \; s^{-1}}} \right)^{-0.38} \left( \sin{\theta} \right)^{-0.28} \; .
\end{aligned}
\end{equation}
The infinitesimal element of solid angle is $\mathrm{d}[\sin(\theta)]$ and so the cumulative probability of encounter up to a given value of theta converges at small angles, and the assumption that most objects impact the surface on normal trajectories is justified, since the average value of $\sin(\theta)$ over the range of impact angles is $2/\pi$, resulting in a correction of order unity in Equation \eqref{finalearthrate} which is disregarded here. An impactor with mass $M_{\mathrm{imp}} \sim 1.7 \times 10^{22} \mathrm{\; g}$ is capable of boiling off all of the oceans on Earth if 100\% of the kinetic energy is converted into thermal energy \citep{2017NatSR...7.5419S}, since the overcoming the vaporization enthalpy of water to vaporize the oceans requires $\sim 3 \times 10^{34} \; \mathrm{erg}$ of energy in addition to the $\sim 3 \times 10^{33} \; \mathrm{erg}$ necessary to heat the water to $100 \; \mathrm{C}$. Figure \ref{fig:earth} shows the Earth impact rate as a function of impactor mass. The chance that life on Earth was sterilized during its lifetime ($\sim 4.5 \mathrm{\; Gyr}$) is $\sim 1 \%$.

\section{Proxima \lowercase{b} sterilization rate}
\label{sec:pbsr}
We interpolate the results in Table 2 of \cite{2018MNRAS.473..295S} to find the dependence of the terrestrial asteroid impact rate on the location of Saturn, in terms of a dimensionless coefficient $\kappa$, for distances of $\sim 3 R_{a, \odot} - 4.5 R_{a, \odot}$, where $R_{a, \odot} \approx 2.7 \mathrm{\; AU}$ is the location of the asteroid belt in the Solar system, normalized to Saturn at a distance of $9.537 \mathrm{\; AU}$ (thereby spanning the range $\kappa \sim 0.26 - 1.25$). We adopt the appropriate scaling for the dimensionless coefficient $\delta$, for the effect of a $\sim 0.1$ Saturn-mass planet on the terrestrial impact rate ($\delta \sim 0.2$), since that is the mass of Proxima c, where $\delta$ is normalized to a Saturn-mass planet given the results in Table 2 of \cite{2018MNRAS.473..295S}. As a caveat, we imagine an architecture similar to the Solar system with a gas giant at location that would allow for a secular resonance near Proxima b.

\begin{figure}
  \centering
  \includegraphics[width=0.9\linewidth]{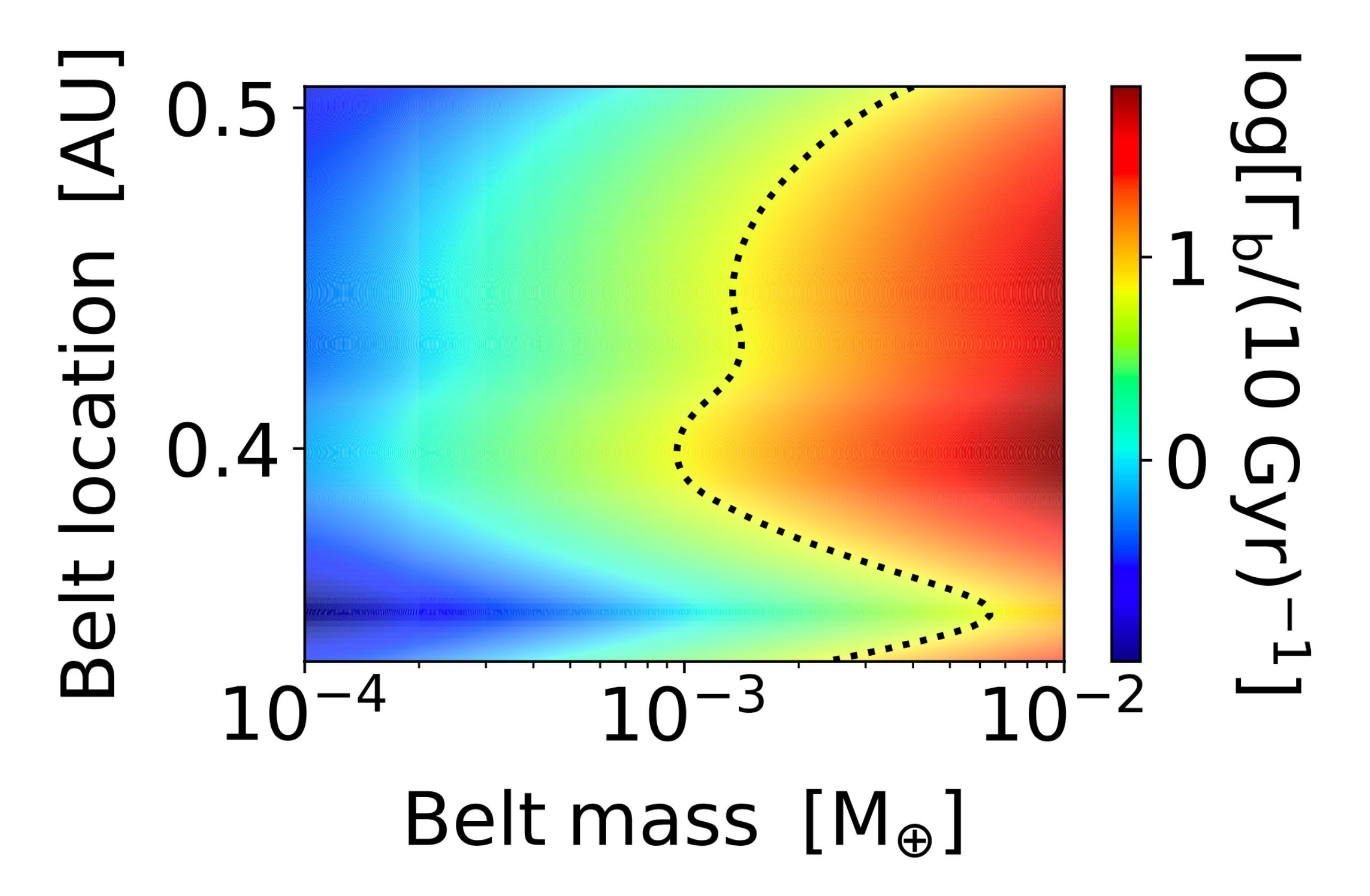}
    \caption{Logarithm of impact rate at Proxima b in units of $\mathrm{(10 \; Gyr)^{-1}}$ as a function of asteroid belt mass in in $M_{\oplus}$ and radius in $\mathrm{AU}$. Proxima c is considered to be a $\sim 0.1$ Saturn-mass outer planet, and the asteroid belt mass is scaled based on that of the solar system, $\sim 4 \times 10^{-4} \mathrm{\; M_{\oplus}}$ \citep{2018AstL...44..554P}. The dotted line indicates an impact with enough energy to boil off all of the oceans on an Earth-like planet, in which 100\% of the kinetic energy is converted into thermal energy.
}
    \label{fig:proxima_b}
\end{figure}

The impact rate is also linearly dependent on the mass of the asteroid belt $M_a$, assuming the size distribution is similar to that of the Solar system (motivated by the fact that the Solar system's small body size distribution is the only one measured, in addition to the similarity of the observed size distribution of interstellar objects \citep{2019arXiv190603270S} to the \cite{1969JGR....74.2531D} collisional model), and inversely dependent on the asteroid belt's orbital period $T_a$. Furthermore, the impact rate is proportional to the cross-section of Proxima b's orbit relative to that of the asteroid belt $(R_b/R_a)^2$, and inversely proportional to the square of the orbital distance of Proxima b, $R_b$, to account for the effect of orbital distance on the one-dimensional cross-section of an Earth-like planet at a fixed planetary radius. The factors enumerated above result in the following dependencies for the asteroid impact rate at Proxima b, $\Gamma_{b} \propto \kappa \delta M_a T_a^{-1} (R_b/R_a)^2 R_b^{-2}$. There is no dependence on $R_b$ since the relative orbital cross-section scales as $R_b^2$ while the relative impact cross-section scales as $R_b^{-2}$. When normalized to solar system values, the rate is expressed as,

\begin{equation}
    \label{proxima_rate}
    \Gamma_{b} \sim \Gamma_{\oplus} \kappa \delta \left( \frac{M_{a, P}}{M_{a, \odot}} \right) \left( \frac{M_{P}}{M_{\odot}} \right)^{1/2}  \left( \frac{R_{a, P}}{R_{a, \odot}} \right)^{-7/2} \left( \frac{r_{b}}{r_{\oplus}} \right)^{2} ,
\end{equation}
where $r_b$ and $r_\oplus$ are the radii of Proxima b and of the Earth, respectively. The sterilizing impact rate at Proxima b as a function of asteroid belt mass and asteroid belt location is shown in Figure \ref{fig:proxima_b}. An asteroid belt with a mass of $\gtrsim 10^{-4} \; M_{\oplus}$ could imply as significant likelihood that Proxima b was sterilized in the past. As discussed in Section \ref{sec:jwst}, the upcoming James Webb Space Telescope\footnote{https://www.jwst.nasa.gov/} (JWST) will able to determine the existence of an asteroid belt at the distances of interest, which correspond here to an angular distance of $\sim 0.3 "$. In addition, yet-undetected planets that may orbit Proxima Centauri would affect the impact rate on Proxima b.

\section{Generalized Sterilization}
\label{sec:ire}

\begin{figure}
  \centering
  \includegraphics[width=0.9\linewidth]{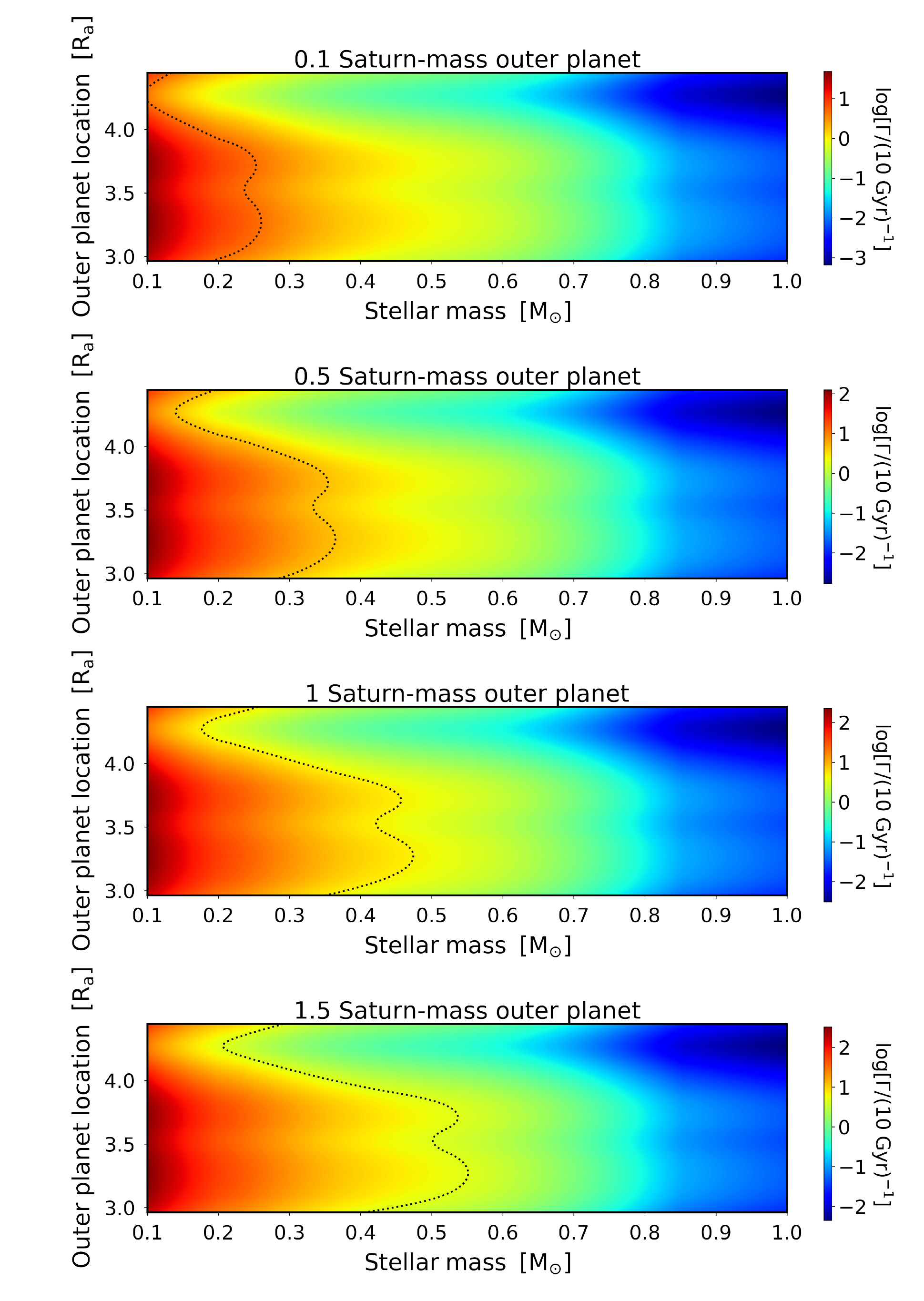}
    \caption{Logarithm of impact rate in units of $\mathrm{(10 \; Gyr)^{-1}}$ as a function of stellar mass in $M_{\odot}$ and outer planet location in units of asteroid belt radius, for different outer planet masses. The dotted lines indicate impacts with enough energy to boil off all of the oceans on an Earth-like planet, in which 100\% of the kinetic energy is converted into thermal energy.
}
    \label{fig:free_parameters}
\end{figure}

The habitable zones around stars scale as, $L_{\star}^{1/2}$. The asteroid belt spans the Solar system's frost line, which separates the terrestrial and giant planets \citep{2006Sci...313.1413R}. To estimate the frost line distances in other planetary system, we fiducially adopt a radiatively heated disk at the time of asteroid belt formation with a simple $L_{\star}^{1/2}$ scaling, which is very similar to the $M_{\star}^{-1/3} L_{\star}^{2/3}$ scaling in \cite{2019A&A...632A...7L}, since the actual scalings for neither frost lines nor asteroid belts around low-mass stars are known \citep{2005ApJ...626.1045I, 2009ApJ...699..824O, 2013MNRAS.428L..11M, 2019A&A...632A...7L}. We estimate the luminosity of stars with masses ranging from $0.2 - 0.85 \; M_{\odot}$ using the mass-luminosity ($M_{\star} - L_{\star}$) relation described in \cite{2018RNAAS...2...19C}, and extrapolate to $\lesssim 0.2 \; M_{\odot}$ and $\gtrsim 0.85 \; M_{\odot}$ with appropriate power-law indices \citep{2003adas.book.....D}. Furthermore, we assume that the asteroid belt mass scales with stellar mass, which is a conservative assumption given that lower-mass stars appear to have higher planet-formation efficiencies \citep{2020AJ....159..247D}. We note that the actual trend of debris mass with stellar mass is yet unknown, as exemplified by the excess debris around $\tau$ Ceti \citep{2004MNRAS.351L..54G}. Adapting Equation \eqref{proxima_rate} to these assumptions of how belt mass and location may scale with stellar mass and luminosity, we find the generalized impact rate to be as follows,

\begin{equation}
    \Gamma \approx \Gamma_{\oplus} \kappa \delta \left( \frac{M_{\star}}{M_{\odot}} \right)^{3/2}  \left( \frac{L_{\star}}{L_{\odot}} \right)^{-7/4} \; \; ,
\end{equation}
which can be coupled with the aforementioned mass-luminosity relation to determine the impact rate on Earth-size planets in habitable zones of their respective stars, as a function of stellar mass. Figure \ref{fig:free_parameters} shows the sterilizing impact rate as a function of stellar mass and outer planet location, for different outer planet masses. We note that the results presented here apply only to a specific two-planet configuration, and that impact rates are highly sensitive to planetary system architecture (see \citealt{2019ESRv..192..445W} for a review).

\section{Discovering outer planets and asteroid belts around other stars}

\label{sec:jwst}
\begin{figure}
  \centering
  \includegraphics[width=0.9\linewidth]{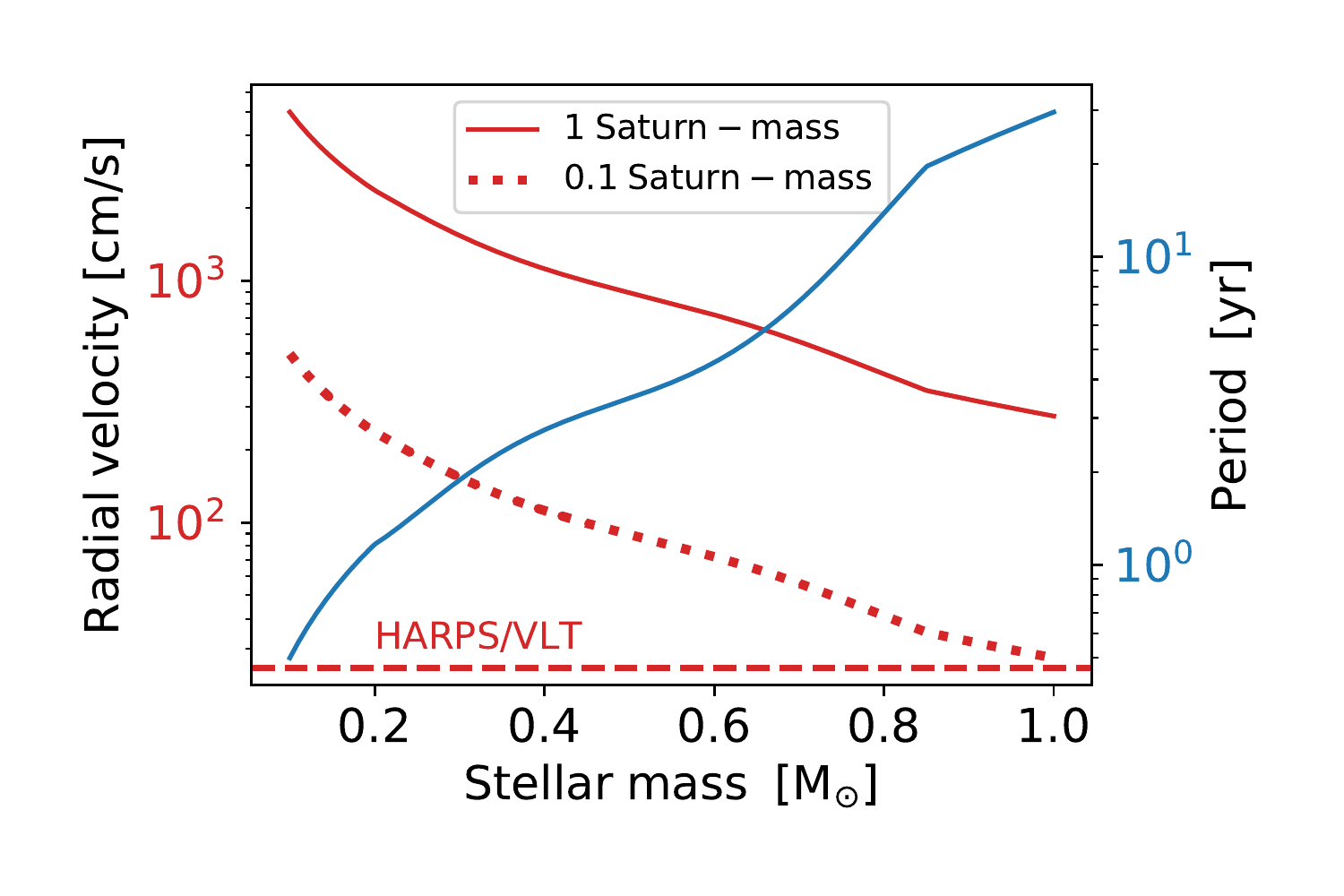}
    \caption{Radial velocity of stars (red) for an edge-on observing geometry in cm/s and orbital period (blue) in years, for Saturn analogs around other stars, adopting the $L_{\star}^{1/2}$ scaling and assuming the distance of Saturn from the Sun. The HARPS/VLT detection limit is shown for reference \citep{2015arXiv150301770P}.
}
    \label{fig:RV}
\end{figure}

For stars with known Earth-like planets in their habitable zones, the existence of an outer planet and an asteroid belt is necessary for finding the sterilizing asteroid impact with the method described here. Outer planets can be discovered by means of radial velocity measurements. For a star of mass $M_{\star}$ and a planet of mass $M_{p}$ at a distance of $R_p$, an edge-on measurement will yield a stellar velocity amplitude of $\pm M_p \sqrt{G/M_{\star} R_p}$. The orbital period of a circular orbit is $\sqrt{4 \pi^2 r^3 / G M_{\star}}$. The results are shown in Figure \ref{fig:RV}, indicating that the existing HARPS/VLT instrument is already capable of discovering planets of the necessary mass and orbital radius around stars of interest. Indeed, this is how Proxima b and c were discovered.

\begin{figure}
  \centering
  \includegraphics[width=0.9\linewidth]{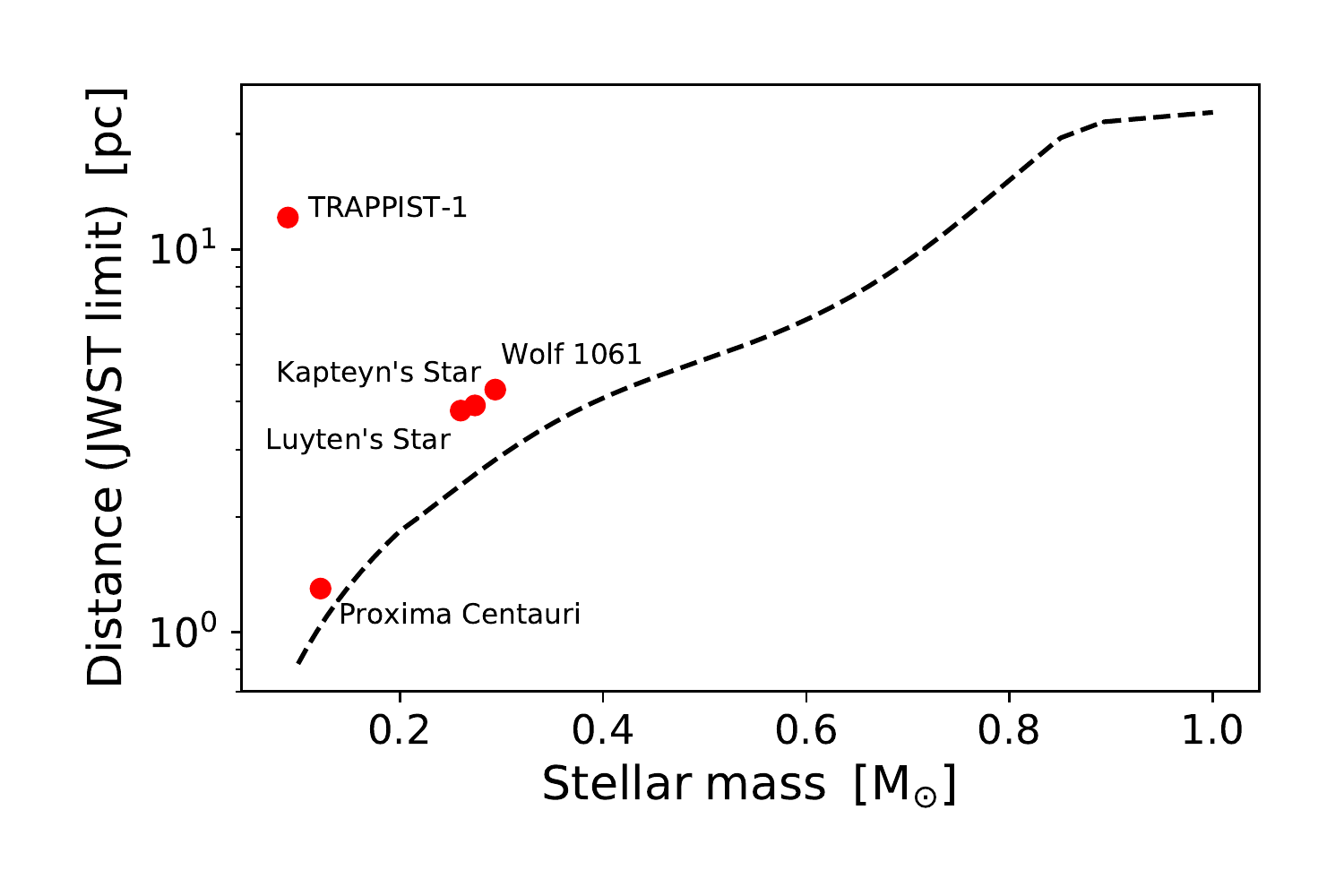}
    \caption{Maximum distance out to which JWST can resolve asteroid belts around other stars, adopting the $L_{\star}^{1/2}$ scaling normalized to the location of the main asteroid belt in the Solar system ($\sim 2.7 \mathrm{\; AU}$). Assuming that the asteroid belt mass scales with stellar mass and that the size distribution is similar to that of the Solar system, above 0.8 AU the maximum distance is flux-limited, as scaled to Spitzer Space Telescope's detection of Epsilon Eridani's debris disk \citep{2009ApJ...690.1522B}, since JWST will observe at the same wavelength. The masses and distances of nearby stars with known potentially habitable exoplanets are displayed for reference.
}
    \label{fig:JWST}
\end{figure}

For detecting evidence of asteroid belts at the distances considered here, we calculate the distance out to which JWST will be able to characterize such belts, given its angular resolution of $\sim 0.1 "$. Figure \ref{fig:JWST} shows the maximum distance as a function of stellar mass, with several stars with known Earth-like planets in their habitable zones plotted for reference. While several stars of interest lie slightly outside the calculated detection limit, we note that the line corresponds to a strict $L_{\star}^{1/2}$ scaling and therefore could underestimate the orbital radii of asteroid belts. Given ALMA's sensitivity to millimeter wavelength emission from a conjectured debris belt of mass $\sim 10^{-3} \; M_{\oplus}$ around Proxima b \citep{2017ApJ...850L...6A}, it is evident that ALMA provides a supplementary probe of debris belts around low-mass stars. In addition, other methods including exoplanet searches can offer independent probes of exo-asteroids \citep{2020arXiv200210370B}.

\newpage

\section{Discussion}
\label{sec:d}

We find that the possible existence of an asteroid belt between Proxima b and Proxima c could result in an existential risk for life on Proxima b due to the expected rate of sterilizing impacts. We imagine an architecture similar to the Solar system with a gas giant at location that would allow for a secular resonance near Proxima b, which can be verified or refuted by future observations.

We generalized the calculation of the sterilizing impact rate to any planetary system with an outer planet and an asteroid belt, allowing for the determination of asteroid impact sterilization risk for habitable exoplanets in constrained planetary architectures. If a debris belt exists around Luyten's star (see Region C described in \citealt{2020arXiv200609403P}), secular resonances with mini-Neptunes GJ 273d or GJ 273e could drive impacts on the potentially habitable planet GJ 273b. The existence of a giant planet \citep{2017AJ....154..103B} and a debris belt \citep{2020MNRAS.492.6067M} exterior to the TRAPPIST-1 planets would affect the sterilization risk in that system.

Future measurements of debris belts with JWST or ALMA and outer planets in systems known to have habitable planets will allow for the magnitude of impact sterilization risk for life to be computed.

\section*{Acknowledgements}
We thank Manasvi Lingam for helpful comments on the manuscript. This work was supported in part by the Origins of Life Summer Undergraduate Research Prize Award and a grant from the Breakthrough Prize Foundation. 

\newpage




\bibliography{bib}{}
\bibliographystyle{aasjournal}



\end{document}